\documentstyle[aps,pra]{revtex}

\begin{document}
\draft
\title{Off-axis vortices in trapped Bose
condensed gases: angular momentum and frequency splitting}
\author{Montserrat Guilleumas and Robert Graham}
\address{Fachbereich Physik, Universit\"at Gesamthochschule Essen.
45117 Essen. Germany}
%
%\date{May 7, 2001}

\maketitle

\begin{abstract}
We consider non centered vortices and their arrays
in a cylindrically trapped
Bose-Einstein condensate at zero temperature.
We study the kinetic energy and
the
angular momentum per particle in
the Thomas Fermi regime and their
dependence on the distance of the vortices from the center of the trap.
Using a perturbative approach with respect to the velocity-field of the
vortices, we calculate to first order the frequency shift of
the collective low-lying excitations due to
the presence of an off-center vortex or a vortex array, and compare
these results with predictions which would be obtained
by the application of
a simple sum-rule approach, previously found to be very successful
for centered vortices. It turns out that the simple sum-rule approach
fails for off-centered vortices.
\end{abstract}
\pacs{03.75.Fi, 05.30.Jp, 32.80.P.j, 67.40.Db}

\section{Introduction}
Vortices in trapped Bose-condensed gases have
recently been observed at JILA \cite{JILA} and at ENS \cite{Paris1,Paris2}.
At ENS they have been
formed by stirring the condensate with a focused laser beam
with an angular frequency $\Omega$.
Experimentally there exists a threshold of the angular frequency of the
stirring beam
($\Omega_c$) to nucleate a single vortex.
When $\Omega \sim \Omega_c$ one vortex is created at the stable
position at the center of the trap. But
at $\Omega > \Omega_c$ and depending on the frequency of the stirring
beam it is possible to create configurations with different number of
vortices forming  vortex arrays
which are stable compared to a single vortex with correspondingly
larger circulation.

The presence of vortices in trapped condensates has been revealed
by time of flight analysis \cite{JILA,Paris1} and recently, by
exciting the quadrupole oscillations of a stirred Bose-Einstein
condensate \cite{Paris2} and observing their frequency splitting. With this
method it has been also possible
to measure the angular momentum induced by the rotation of the trap.
Direct measurement of the angular momentum is difficult in atomic gases,
since most of the diagnostic
techniques provide information on the density profiles.
As expected, for stirring frequencies below $\Omega_c$
no angular momentum is observed \cite{Paris2},
since no vortex can be nucleated, and
just above $\Omega_c$ the angular momentum has a jump of
$\sim \hbar$ which indicates
the presence of a stable singly quantized vortex
at the center of the trap.
However, for higher stirring frequencies,
an apriori unexpected behavior of the angular momentum has been
experimentally found \cite{Paris2}: for $\Omega > \Omega_c$
when vortex arrays are
nucleated,
the measured angular momentum does
not show any jump of size $\hbar$  as a new singly quantized vortex is
nucleated, as is e.g. observed for the corresponding experiment
in superfluid Helium \cite{Donnelly}; instead it is a
smooth and increasing function of $\Omega$.
We shall see that this is not in contradiction with the quantization
of the circulation of a vortex, but it is a consequence of the nucleation
of off-axis vortices that form a vortex array.

The purpose of the present paper is to provide a detailed analysis
of the dependence of the kinetic energy and the angular momentum on
the distance of an
off-centered vortex with respect to the symmetry axis.
We will
generalize the results to vortex arrays to give a physical insight
of the smooth behavior of the angular momentum found experimentally.

In an axially symmetric trap the axial component of angular momentum
($m$) is a good quantum number, and in the absence of vortices
the collective excitations of the condensate are degenerate with respect
to the sign of $m$.
However, the presence of vortices breaks this degeneracy, and
elementary excitations carrying opposite angular
momentum are no longer degenerate.
The frequency shifts of the quadrupole oscillations due to the
presence of a
vortex on the symmetry-axis of the trap have been calculated in
\cite{sumrule} using a sum rule approach.
Analytic results for the energy splitting are obtained in \cite{sumrule}
in the Thomas-Fermi limit
%(large number of condensate atoms and positive scattering length)
for the quadrupole modes
and have been used
in \cite{Paris2} to measure the angular
momentum of the condensate.
General expressions based on the hydrodynamic approach
for the energy splitting due to a centered vortex
have been obtained in \cite{Sinha97,Svidzinsky98}
within perturbation theory.

In this paper we consider off-centered quantized vortex lines
in large cylindrically confined condensates at zero temperature,
and therefore, no dissipation mechanism \cite{Fedichev00} is
taken into account.
We calculate the kinetic energy and the angular momentum with respect to the
symmetry axis \cite{Fetter65,Hess67} of an array of
vortex lines in a trapped Bose condensed gas.
Using the perturbative  approach proposed in \cite{Svidzinsky98} and assuming the
Thomas-Fermi limit,
we generalize the analytic expressions for the energy splitting of the
low-lying collective modes to consider the effect of the distance of
off-centered vortices
with respect to the symmetry axis. We also compare with the results obtained
by a direct extension of the sum-rule approach to off-center vortices, but find that
such an extension, at least in the direct version examined here, fails.

\section{Vortex states}
We consider a weakly interacting Bose-condensed gas confined
in a harmonic trap $V_{\rm ext}({\bf r})$ at zero temperature.
The condensate wave function can be written in terms of its density
$\rho({\bf r})=|\Psi({\bf r})|^2$ and phase $S({\bf r})$ by
\begin{equation}
\Psi({\bf r})=\sqrt{\rho({\bf r})} exp[i S({\bf r})] \,.
\label{wf}
\end{equation}
The number of atoms in the condensate is
$\int \! d{\bf r} |\Psi|^2= N$, and the
superfluid velocity is given by ${\bf v}=(\hbar/M) {\bf \nabla} S$,
where $M$ is the atomic mass.
The ground state wave function in the absence of vortices has
a spatially constant phase and therefore
zero velocity,
but when a quantized vortex is present the phase and the velocity
field of the ground state have to be determined according to the
boundary conditions of the system.

\subsection{Condensate Density}
The ground state wave function $\Psi({\bf r})$ is a stationary solution
of the Gross-Pitaevskii equation
\begin{equation}
\left( - { \hbar^2 \nabla^2 \over 2M} + V_{\rm ext}({\bf r})
+ g \mid \!\Psi({\bf r}) \!\mid^2 \right) \Psi({\bf r}) =
\mu \Psi({\bf r}) \; ,
\label{GP}
\end{equation}
where the coupling constant $g$ is given by
the  $s$-wave scattering length $a$ through
$g=4 \pi \hbar^2 a/M$, and  $\mu$ is the chemical potential fixed
by the normalization condition of the ground state.

Non-rotating experimental traps \cite{JILA,Paris1,Paris2}
have axial symmetry, with different radial
($\omega_{\perp}$) and axial ($\omega_z$) trapping frequencies,
whose ratio defines the anisotropy parameter
$\lambda = \omega_z/\omega_\perp$.
So far, vortex arrays have been produced at ENS \cite{Paris1,Paris2}
in a highly anisotropic cigar-shaped trap
(quasi-cylindrically symmetric) with $\lambda \sim 0.05$.
Thus, for simplicity,
we consider an idealized cylindrical trap that is
uniform in the $z$-direction ($\omega_z=0$) and has a
harmonic confining potential in the radial direction of the form
\begin{equation}
V_{\rm ext}(r_\perp)=\frac{1}{2} M \omega_{\perp}^2 r_\perp^2 \,,
\label{Vtrap}
\end{equation}
with $r_\perp^2=x^2+y^2$.
The harmonic trap frequency $\omega_\perp$ provides a typical length scale
for the system,
$a_{\perp}=(\hbar/M \omega_{\perp})^{1/2}$.

We consider the Thomas-Fermi (TF) regime \cite{Trento},
valid for large condensates with positive scattering length, where
the kinetic pressure can be neglected compared to the
interaction energy-density.
The ground state density of the condensate,
in the absence of vortices, is given by
\begin{equation}
\rho_0(r_{\perp})= \frac{\mu}{g}\left(1-\frac{r_{\perp}^2}
     {R_{\perp}^2}\right) \,,
\label{TFdensity}
\end{equation}
for $r_\perp \leq R_\perp$, and $\rho_0(r_{\perp})=0$ elsewhere. The
Thomas-Fermi radius of the cylindrical condensate is
$R_\perp=(2 \mu / M \omega_{\perp}^2)^{1/2}$. The density is defined
in the interval $-R_z \leq z \leq R_z$, where $2 R_z$ is the length of the
cylinder. In the cylindrical geometry the
validity of the Thomas-Fermi approximation for the ground state is
guaranteed by the condition $N a/(2 R_z) = N_\perp a\gg 1$,
or equivalently, $\mu \gg \hbar \omega_\perp$,
where $N_\perp=N/(2 R_z)$ is the number of atoms per unit length
\cite{Lundh00}. From Eq.~(\ref{TFdensity})
\begin{equation}
N_\perp=\int_0^{R_\perp} \rho_0 \, 2 \pi r_\perp dr_\perp=
          \frac{\mu}{g}\frac{\pi R_\perp^2}{2}\,.
\label{N/Rz}
\end{equation}

When a quantized vortex is present
at the position ${\bf r}_{\perp 0}$,
the density of the system
% tends to zero when ${\bf r} \rightarrow {\bf r}_0$
drops to zero at the center of the vortex core
whose size is determined by the healing length $\xi$.
In the limit of large systems it can be approximated by
$\xi=(8 \pi \rho_0({\bf r}_{\perp 0}) a)^{-1/2}$,
where $\rho_0({\bf r}_{\perp 0})$ is the density of the condensate at the
position of the vortex core but in the absence of vorticity.
%center of the trap in absence of vorticity.
For a centered vortex,
Eq.~(\ref{TFdensity}) yields $\rho_0(0)=\mu/g$, and the corresponding
healing length $\xi_0=\xi(0)$ can be rewritten as
\begin{equation}
\frac{\xi_0}{R_\perp}=\left(\frac{a_\perp}{R_\perp}\right)^2 \,,
\label{healing}
\end{equation}
implying the following set of TF inequalities
for the length scales
$\xi_0 \ll a_\perp \ll R_\perp$.
Analogously, from (\ref{TFdensity}) the local healing length
\cite{Lundh00} is
\begin{equation}
\xi(r_{\perp 0}/R_\perp)=\frac{\xi_0}
{\sqrt{1-r_{\perp 0}^2/R_\perp^2}} \,.
\label{localhealing}
\end{equation}
Within the TF approximation, analytical expressions for the
density of a vortex state have been obtained for a centered vortex
\cite{Sinha97,Svidzinsky98} and for a straight off-axis vortex line
\cite{Jackson99}. However,
in weighted spatial averages of quantities varying only on the scale
$R_\perp$, the density of a vortex state may be replaced by the
density of the vortex-free state, the corrections being only
of order
$(\xi/R_\perp)^2=(a_\perp/R_\perp)^4$ \cite{Svidzinsky98}, which
is negligible.

\subsection{Velocity Field}
We consider states having a quantized vortex line along the $z$-axis
and all the atoms flowing around it with quantized circulation.
The property of a single-valued wave function leads to the quantization
of the circulation around an arbitrary closed loop which encloses the
vortex core
\begin{equation}
{\cal K}=\oint {\bf v} \cdot d{\bf l} =
   \frac{\hbar}{M} \oint {\bbox \nabla} S \cdot d{\bf l}=\frac{h \kappa}{M} \,,
\label{circulation}
\end{equation}
where the integer number $\kappa$ is the quantum of circulation.
For sufficiently large frequencies $\Omega > \Omega_c$ a single vortex line
with an integer
quantum number $\kappa > 1$ can, in principle, appear but this state is unstable
and fragments into a vortex array formed by
$\kappa$ vortices each with a unit of circulation and
position ${\bf r}_{\perp 0 i}$ ($1\le i\le\kappa$)
relative to the symmetry axis.

We will not consider large arrays of vortex lines, where an average
vortex density can be defined \cite{Donnelly,Fedichev00,Fetter65},
but small ones, corresponding to the
experimental region \cite{Paris1,Paris2} of multiple vortices
($\kappa \leq 5 $). Large
vortex arrays may enter the "turbulent" region found experimentally.
For small vortex arrays the density of the system can be approximated
by the density of a vortex-free condensate, because
vortices rotating in the same direction experience an effective
repulsive interaction \cite{Donnelly},
and the inhomogeneities due to the well separated
vortex cores are negligible in the condensate density
($\xi \ll R_\perp$).

Let us assume a generic quantized vortex line parallel to
the $z$-axis at
the position ${\bf r}_{\perp 0}$ and with quantized circulation $\kappa$.
Then, from Eq.~(\ref{circulation}),
${\bf \nabla} \times {\bf v}= (h \kappa/M)
      \delta({\bf r}_\perp-{\bf r}_{\perp 0}) \hat{z}$,
where $\hat{z}$ is the unit vector in the $z$-direction,
and the superfluid is irrotational everywhere except for the vortex
core at ${\bf r}_\perp={\bf r}_{\perp 0}$, where the density vanishes.

For a positively oriented vortex in an infinite, uniform
system, the velocity then is
\begin{equation}
{\bf v}_{uni}({\bf r}_\perp)= \frac{\hbar \kappa}{M}
\frac{\hat{z}\times ({\bf r}_\perp-{\bf r}_{\perp 0})}{|{\bf r}_\perp-{\bf
r}_{\perp 0}|^2} \,.
\label{vuniform}
\end{equation}
In particular, for a centered vortex only the tangential component
carries non vanishing atomic flux, and
${\bf v}=\hbar \kappa/(M r_\perp) \hat{\varphi}$, where
$\hat{\varphi}$ is the unit vector in the tangential direction
in cylindrical coordinates $(r_\perp,\varphi,z)$. Eq.~(\ref{vuniform})
once again shows that a cut-off of the order of the local healing length
$\xi$ for the distance $|{\bf r}_\perp -{\bf r}_{\perp 0}|$ is needed,
because at distances smaller than $\sqrt{2}\xi$ the velocity (\ref{vuniform})  
surpasses the local velocity of sound $c_s({\bf r}_\perp)=\sqrt{g|\psi({\bf
r}_\perp)|^2/M}$.

For a confined system, the velocity field is affected by the
boundary of the system and by the spatially varying density, and has to
fulfill the following physical conditions:
i) the normal velocity has to vanish at the boundary,
and ii) the condition for
stationary flow ${\bbox \nabla}\! \cdot\! (\rho {\bf v}) =0$ has to be
satisfied.
It is well known from the rotating bucket experiment in superfluid helium
\cite{Fetter65,Hess67,Lundh00} that for an homogeneous system
confined in a cylinder of radius $R_\perp$,
the normal velocity vanishes at the
boundary by introducing an oppositely oriented image vortex at
${\bf r}_{\perp 1}= (R_\perp/r_{\perp 0})^2 \,{\bf r}_{\perp 0}$.
The resulting
velocity field is
\begin{equation}
{\bf v}_{0}({\bf r}_\perp)= \frac{\hbar \kappa}{M} \frac{\hat{z} \times
({\bf r}_\perp-{\bf r}_{\perp 0})}
   {|{\bf r}_\perp-{\bf r}_{\perp 0}|^2} - \frac{\hbar \kappa}{M} \frac{\hat{z}
       \times ({\bf r}_\perp-{\bf r}_{\perp 1})}{|{\bf r}_\perp-{\bf
r}_{\perp 1}|^2} \,.
\label{vimage}
\end{equation}
If the system has a density gradient, the condition
ii) for stationary flow
is fulfilled by introducing a small correction to the velocity which can
be neglected when the density varies over a larger scale than the healing
length \cite{Lundh00}. Thus, in the TF limit the velocity field can be
approximated as ${\bf v} \simeq {\bf v}_0$.

A vortex at ${\bf r}_{\perp 0}$ is influenced by the velocity-field induced
by its mirror-vortex. If ${\bf r}_{\perp 0}$ approaches the boundary  this
velocity diverges, signalling the break-down of the Thomas-Fermi
approximation. A suitable lower cut-off for the
distance from the boundary, $\xi_{boundary}=R_\perp(\xi_0/2R_\perp)^{2/3}$,
is implied by restricting the velocities induced by the mirror-vortices by
the local velocity of sound.

%In the absence of dissipation the vortex line at ${\bf r}_{\perp 0}$ at i
%a distance
%$r_{\perp 0}$ from the $z$-axis moves with the velocity created there by the  
%image-vortex. This causes  a precession of the vortex line around the
%$z$-axis
%with angular velocity
%\begin{equation}
%\Omega_1=\frac{\hbar}{M}\frac{1}{R_\perp^2-r_{\perp 0}^2}.
%\end{equation}

From Eq.~(\ref{vimage}) one can see that the contribution of a vortex
(vortex and image vortex) is additive. Therefore, the velocity field
corresponding to a vortex array formed by $\kappa$ singly
quantized vortices at ${\bf r}_{\perp 0 i}$ with $i=1...\kappa$ is then
\begin{equation}
{\bf v}({\bf r}_\perp)= \sum_{i=1}^\kappa{\bf v}_i({\bf r}_\perp)
\label{varray0}
\end{equation}
with
\begin{equation}
{\bf v}_i({\bf r}_\perp)=  \left(\frac{\hbar}{M} \frac{\hat{z}
   \times ({\bf r}_\perp-{\bf r}_{\perp 0 i})}
   {|{\bf r}_\perp-{\bf r}_{\perp 0 i}|^2} - \frac{\hbar}{M} \frac{\hat{z}
   \times ({\bf r}_\perp-{\bf r}_{\perp 1 i})}{|{\bf r}_\perp-{\bf r}_{\perp  
1 i}|^2}\right) \,,
\label{varray}
\end{equation}
where ${\bf r}_{\perp 1 i}= (R_\perp/r_{\perp 0 i})^2 \,{\bf r}_{\perp 0 i}$
is the position of the vortex image corresponding to the $i$-vortex.
%Each vortex now moves in the velocity field created by all the other vortices  
%and all image vortices. For symmetrical arrays of $2, 3$, or $4$ vortices at  
%equal distances $r_{\perp 0}$ from the $z$-axis this leads to a rigid
%rotation
%of the arrays around the $z$-axis with angular frequencies
%\begin{eqnarray}
%&&\Omega_2=\frac{\hbar}{M}\frac{1}{2r_{\perp 0}^2}\frac{3r_{\perp
%0}^4+R_\perp^4}{R_\perp^4-r_{\perp 0}^4}\nonumber \\
%&&\Omega_3=\frac{\hbar}{M}\frac{1}{r_{\perp 0}^2}\frac{2r_{\perp
%0}^6+R_\perp^6}{(R_\perp^2-r_{\perp 0}^2)(R_\perp^4+R_\perp^2r_{\perp
%0}^2+r_{\perp 0}^4)}\\
%&&\Omega_4=\frac{\hbar}{M}\frac{1}{2r_{\perp 0}^2}\frac{5r_{\perp
%0}^8+3R_\perp^8}{R_\perp^8-r_{\perp 0}^8}\nonumber,
%\end{eqnarray}
%where $\Omega_1<\Omega_2<\Omega_3<\Omega_4.$
%If ${\bf r}_{\perp 0}$ approaches the boundary the $\Omega_i$ and the
%corresponding rotation velocity diverge. A suitable lower cut-off for the
%distance from the boundary, $\xi_{boundary}=R_\perp(\xi_0/2R_\perp)^{2/3}$,
%is implied by restricting the velocities by the local velocity of sound.

\section{Kinetic Energy}
In the Thomas-Fermi limit the dominant part of the excess energy of a vortex  
state over the ground state without a vortex is given by the kinetic energy of  
the velocity field \cite{Sinha97}. It is obtained from
(\ref{varray0},\ref{varray}) in the
form
\begin{equation}
E_{kin}=\sum_{i=1}^\kappa E_i +\sum_{i=1}^\kappa\sum_{j=1}^{i-1}E_{ij}
\label{E1}
\end{equation}
with
\begin{eqnarray}
E_i =\frac{M}{2}&&\int_{-R_z}^{R_z}dz\int_{0}^{R_\perp}dr_\perp
r_\perp\rho_0(r_\perp,z)\nonumber\\
&&\int_0^{2\pi}d\varphi\, {\bf v}_i^2(r_\perp,z,\varphi)
\label{Ekin1}\\
E_{ij} =M&&\int_{-R_z}^{R_z}dz\int_{0}^{R_\perp}dr_\perp
r_\perp\rho_0(r_\perp,z)\nonumber\\
&&\int_0^{2\pi}d\varphi\, {\bf v}_i(r_\perp,z,\varphi)\cdot{\bf
v}_j(r_\perp,z,\varphi) \,,\label{Ekin2}
\end{eqnarray}
where $E_i$ is the self energy of the $i$-vortex and
$E_{ij}$ the binary interaction energy between a couple of vortices
$i$ and $j$.
Let us define the dimensionless quantities
${\tilde E}_i={E_i}/{\left(\frac{2N\hbar^2}{MR_\perp^2}\right)}$,
and analogously ${\tilde E}_{ij}$ and ${\tilde E}_{kin}$.
For an array of vortex lines along the $z$-axis in the
idealized cylindrical trap described by Eqs.~(\ref{Vtrap},\ref{TFdensity}), 
the integrals in
(\ref{Ekin1},\ref{Ekin2}) can be performed and we obtain after some
calculations\footnote{It is useful for this purpose to adopt a complex
representation of the transverse vector ${\bf r}_\perp\rightarrow z=x+iy$ and  
of the velocity field of a vortex at $z_0$ in the form ${\bf v}_j \rightarrow  
v_j=(\hbar/M)iz/(r_\perp^2-zz^*_0)$ with complex
conjugate $v^*_j=-(\hbar/M)i/(z-z_0)$ and to perform the integral
over $\varphi$ at fixed value of $r_\perp=|z|$ as a complex
contour-integration making use of the residuum-theorem.}, defining the
rescaled distance-vector of the vortex-line from the $z$-axis
${\bf x}_i={\bf r}_{\perp 0 i}/R_\perp$,
\begin{eqnarray}
%\frac{E_i}{\left(\frac{2N\hbar^2}{MR_\perp^2}\right)}=
{\tilde E_i}=
&&\frac{1-x_i^4}{2x_i^2}\ln(1-x_i^2)+(1-x_i^2)\ln\left(\frac{R_\perp}
{\sqrt{2}\xi_0}\right)\label{E4}\\
%\frac{E_{ij}}{\left(\frac{2N\hbar^2}{MR_\perp^2}\right)}=
{\tilde E_{ij}}=
&&(1-{\bf
x}_i\cdot{\bf x}_j)\ln\left(\frac{1}{({\bf x}_i-{\bf
x}_j)^2}\right)\nonumber\\
&&+\left(1+\frac{1-x_i^2-x_j^2-x_i^2x_j^2}{2x_i^2x_j^2}{\bf x}_i\cdot{\bf
x}_j\right)\nonumber\\
&&~~~\ln(1+x_i^2x_j^2-2{\bf x}_i\cdot{\bf x}_j)\nonumber\\
\label{E5}\\
&&-\frac{1-x_i^2-x_j^2+x_i^2x_j^2}{x_i^2x_j^2}|{\bf x}_i\times{\bf x}_j|\nonumber\\
&& ~~~\arctan\left(\frac{|{\bf x}_i\times{\bf x}_j|}{1-{\bf x}_i\cdot{\bf
x}_j}\right).\nonumber
\end{eqnarray}
Here the healing length $\sqrt{2}\xi_0/R_\perp$ is used as a cut-off. It is of  
interest to
remark that for $R_\perp$ and $\xi_0$ fixed, both $E_i$ and $E_{ij}$ are
proportional
to $N$. The expression (\ref{E4}) for the kinetic energy of a vortex in the
presence of its image-vortex (self energy)
agrees with the result presented by Fetter \cite{fetter} for one vortex.
The kinetic energy of
an array of vortices without their images was calculated by Castin and Dum
\cite{Castin99}. Due to the presence of image vortices,
introduced to ensure the vanishing of the normal component of the
velocity at the boundary,
our result differs from that in \cite{Castin99}
but it is only slightly more complicated.

In Figure 1 we plot the kinetic energy of symmetrical arrays of
singly quantized
1,~2,~3,~4 vortices, analogous to the experimental configurations of
Ref.~\cite{Paris2},
as a function of the normalized
distance from the trap axis. The kinetic energy goes to zero as the distance  
of the vortices from the boundary of the condensate goes to zero. This is a
consequence of the boundary condition on the surface, where each vortex meets  
its image and thus is annihilated in the process. As mentioned in
the preceding section a lower cut-off must be imposed on the distance of the  
vortex-line from the boundary. This is tantamount to restricting $x_i$ by $x_i  
< 1-(\xi_0/2R_\perp)^{2/3}$. The energy is maximal at the center of the trap,  
where the expression (\ref{E5}) diverges logarithmically unless a cut-off on
the minimal distance of the vortex-cores from the $z$-axis is imposed, whose
size is again taken as $\sqrt{2}\xi_0/R_\perp$.
Fig.~1 makes it clear that a single vortex or a vortex array must be created at
the boundary where the energy required to set up the necessary velocity field
becomes vanishingly small. However, as we have already mentioned,
the description of a vortex close to
the boundary is outside the scope of the present approach, because there the  
Thomas-Fermi approximation, on which it is based, breaks down.

\section{Angular Momentum}
Let us recall the expression of the angular momentum of a vortex
state in an imperfect Bose gas \cite{Fetter65,Hess67}. We
consider an axially symmetric condensate and are interested in the angular
momentum around
its symmetry axis defined along the $z$-axis.
For a single vortex at a distance ${r}_{\perp 0}$ from the symmetry axis,
the angular momentum of the system
with respect to the $z$-axis must be calculated by integrating
over the whole volume of the system:
\begin{equation}
<{\bf L}>= \int d^3{\bf r}\,\rho({\bf r})\,({\bf r} \times {\bf v})\,,
\label{L}
\end{equation}
where ${\bf v}$ is the superfluid velocity around the vortex and
$\rho({\bf r})$ is the condensate density.
As discussed already, we can neglect in the TF limit the
effect of the vortex on the density profile and take
$\rho({\bf r})\simeq\rho_0({\bf r})$.
Then, using cylindrical coordinates around the $z$-axis,
we have
$\rho_0({\bf r})=\rho_0(r_\perp,z)$.
In the present geometry, $\bf L$ is parallel to the axis of rotation
and its magnitude is given by
\begin{equation}
<L_z>= \int d^3{\bf r} \, \rho_0(r_{\perp},z) \hat{z}
   \cdot ({\bf r} \times {\bf v})\,.
\label{Lz}
\end{equation}
The triple scalar product can be written as
$\hat{z} \cdot ({\bf r} \times {\bf v})=
       r_{\perp} \, v_{\varphi}$,
where $v_{\varphi}$ is the tangential component of the velocity.
Assuming that the boundaries of the system are given by
$-R_z \leq z \leq R_z$, $0 \leq r_{\perp} \leq R_{\perp}$,
then
\begin{equation}
<L_z>= \int_{-R_z}^{R_z} dz \,
     \int_{0}^{R_{\perp}} \rho_0(r_{\perp},z) \, r_{\perp} \,
  dr_{\perp} \int_{0}^{2 \pi} v_{\varphi} \,r_\perp \,d{\varphi}\,.
\label{Lz1}
\end{equation}
It can be easily seen that, for fixed $r_\perp$ and $z$, the
angular integral is a line integral around a closed path,
which corresponds to the quantization of the circulation
(\ref{circulation}).
Therefore, the angular integral in (\ref{Lz1}) will
contribute only when the closed contour
encloses the vortex core.
That is,
\begin{equation}
\int_{0}^{2 \pi} v_{\varphi}\, r_\perp d{\varphi} =
  \oint_{r_{\perp},z} {\bf v} \, \cdot \, d{\bf l} =
  \frac{\hbar \kappa}{m} 2 \pi \, \theta(r_\perp-r_{\perp 0}) \,,
\label{contour}
\end{equation}
where $\theta(r_\perp-r_{\perp 0})$ is the step function.
Remarkably, the detailed form of the velocity field around the vortex
drops out in this expression. In particular, the mirror vortex
contributing in (\ref{varray}) does not contribute to this integral
at all because it is positioned outside the condensate and therefore
never enclosed by the integration contour in (\ref{contour}).
This leads to the following result for the angular momentum
\begin{equation}
<L_z>=\int_{-R_z}^{R_z} dz \, \int_{r_{\perp 0}}^{R_\perp}
  \frac{2 \pi \hbar \kappa}{m}\,r_{\perp} \,
  \rho_0(r_{\perp},z)\,dr_{\perp} \,.
\label{Lz2}
\end{equation}
Due to the step function, only the part of the Bose condensate
at $r_\perp \geq r_{\perp 0}$, that is outside the
smallest circle around the $z$-axis which still encloses the
vortex-line, contributes
to the total angular momentum and therefore to the moment
of inertia of the condensate.\footnote{The 
expression (\ref{Lz2}) can be generalized to
take into account a more realistic
three dimensional vortex line.
Real axially symmetric traps are not cylindrical,
but provide also a longitudinal harmonic confinement ($\omega_z \neq 0$).
It induces an inhomogeneity
and a density gradient of the condensate along the longitudinal
direction which will affect the velocity.
The resulting vortex line must then deform along its length
\cite{Castin99,Svidzinsky00} and meet the boundary of the condensate at a
right angle in order to satisfy
the physical conditions for the velocity field. Therefore,
the distance vector of a quantized vortex from the $z$-axis
in general depends on $z$: ${r}_{\perp 0}={r}_{\perp 0}(z)$.
By taking as a limit of radial integration $r_{\perp 0}(z)$
in (\ref{Lz2}),
the angular momentum expression (\ref{Lz2}) allows to take
into account such more complicated vortex configurations.}

We will consider a large cylindrical condensate,
which provides a good approximation for very elongated
cigar-shaped traps in the Thomas-Fermi regime. Inserting in
Eq.~(\ref{Lz2}) the Thomas-Fermi density (\ref{TFdensity}) it follows
that the angular momentum per particle is
\begin{equation}
<l_z>= \frac{<L_z>}{N}=
\hbar\kappa\,\left[1-\left(\frac{r_{\perp 0}}{R_\perp}\right)^2\right]^2 \,.
\label{Lz/N}
\end{equation}
For fixed $R_\perp$ it is independent of $N$. It is worth stressing that
Eq.~(\ref{Lz/N}) is the dominant term
of the angular momentum, where corrections of order
$(\xi/R_\perp)^2=(a_\perp/R_\perp)^4$ are neglected.
Eq.~(\ref{Lz/N}) shows that even though the circulation
(\ref{circulation}) around each vortex is quantized in units of
$\hbar/M$ the angular momentum per particle around the $z$-axis is
not quantized in units of $\hbar$, in general.

Since $\bf L$ is linear in the velocity field, contributions from
additional vortices are additive and the angular momentum per
particle (\ref{Lz/N}) can be generalized to a vortex array
of $\kappa$ singly quantized vortices at
${\bf r}_{\perp 0 i}$ ($i=1...\kappa$) as
\begin{equation}
<l_z>= \hbar \sum_{i=1}^{\kappa}
         \left[1-\left(\frac{r_{\perp 0 i}}{R_\perp}\right)^2\right]^2 \,.
\label{Lz/Narray}
\end{equation}
It is interesting to note that for a singly quantized
vortex ($\kappa=1$), $<l_z>$ is equal to $\hbar$ only when
the vortex line is centered at the $z$-axis. Otherwise $<l_z>$
is strictly less than $\hbar$ and decreases when the position
of the vortex core moves away from the center and
approaches the boundary of the condensate. When the vortex
core reaches the edge of the condensate ($r_{\perp 0} \simeq R_\perp$)
then $<l_z> \simeq 0$.
Analogously, when $\kappa$ singly quantized vortices are present,
the angular momentum per particle is lower than $\kappa \hbar$
unless all cores are along the $z$-axis \cite{Paris2}.
But this configuration corresponds to one vortex with
circulation $\kappa \hbar/M$ which is unstable when $\kappa >1$
\cite{unstable}
and breaks into an array of $\kappa$ vortices all with unit
quantization $\hbar/M$.

Let us now calculate the angular momentum per particle in the simplest
vortex-arrays in cylindrical traps rotating with angular
velocity $\Omega$. They are created
if the array is permitted to reach a state of relative
equilibrium minimizing the energy
\begin{equation}
E_{kin}(\Omega)=E_{kin}(0)-\Omega<L_z>.\label{E6}
\end{equation}
Here $E_{kin}(0)$ is the kinetic energy (\ref{E1}) of the vortex-array
in the non-rotating trap.
As before we consider symmetrical arrays of 1-4 vortices.
In Figure 2 the kinetic energy (\ref{E6}) is plotted against the common
normalized
distance of the vortices from the $z$-axis, for an angular
velocity $\Omega/2\pi=
40$ Hz, which is, in our idealized 2-dimensional trap and within the
Thomas-Fermi approximation, the critical frequency where the 2-vortex array
first becomes stable. The corresponding critical frequencies for the
symmetrical 1,
3, and 4 vortex arrays are, respectively $\Omega_c= 33.6, 43.3$, and $46$ Hz,  
where we have chosen for the sake of concreteness, the values of $M, R_\perp$  
according to the
experiment described in \cite{Paris2}.
As expected, the critical frequency is larger for large vortex arrays.
These values are different from and much lower than the measured critical
rotation frequencies at which vortices or vortex arrays are first observed
to appear, because of the existence of energy-barriers which must be overcome  
before the relative energy-minima formed by the vortex states can be reached.

It can be seen that in the rotating trap, for
sufficiently high values of $\Omega$, a single vortex is in relative or
absolute equilibrium only at the rotation axis as is of course well-known,
while the symmetrical arrays of vortices have equilibria at finite distances  
from the $z$-axis which increase with the number of vortices because of their
mutual repulsion. In Figure 3 we plot for the same vortex-arrays
at the corresponding equilibrium configuration
the average angular-momentum per particle as a function of $\Omega$. With
increasing $\Omega$ the equilibrium-positions of the vortices in the arrays
move towards the $z$-axis and the average angular-momentum therefore increases.
For a single vortex, since the stable position is at the trap center
independently of the rotation frequency, the angular momentum is
always $1 \hbar$.

Let us remark here that, in the absence of dissipation, vortices created
experimentally in
rotating traps, in general, need
not  correspond to minima of (\ref{E6}). A dissipative mechanism must be
active on the scale of the time-interval during which the rotation-frequency  
is switched on for their positions to be able to relax to an energy minimum.  
After the rotation is switched off the same mechanism will tend to lead
to a relaxation of $r_{\perp 0}$ towards the minimum at the boundary
$r_{\perp 0}=R_\perp$ of the energy (\ref{E1}), which increases the observed  
value of $r_{\perp 0}$.
In the experiments on vortices in traps reported so far dissipation seems to  
play a negligible role. In the absence of such a mechanism,
however, the vortices
with distances $r_{\perp 0}$ outside the energy-minimum cannot relax but
experience a force and a corresponding Magnus deflection already in the
rotating trap which leads to
a rotation of an array of equidistant vortices around the $z$-axis even
in the frame
rotating with the trap.
Vortices which are created at the boundary and have not yet reached their
equilibrium-distance from the $z$-axis will
then be observed to  have an angular momentum which is smaller than the
equilibrium value shown in Fig.~3.
From this point of view, the fact that the first vortex which is formed is
{\it always} observed to be in the center seems to indicate that the first
vortex
is, in fact, not created by the motion of a vortex-line from the boundary to  
the center but by a different mechanism, like e.g. the condensation of
collective excitations with $l=1$ into a vortex state with $l=1$.

Let us now turn to a discussion of some related experimental results
which have been obtained in Refs.\cite{Paris1,Paris2}. From the
transverse absorption images of Ref.~\cite{Paris1} we
estimate the angular momentum corresponding to the experimental
configurations of the condensate with $1$ up to $4$ vortices.
We proceed as in \cite{Paris2}, first
we obtain a qualitative measure of the ratio between the
distance of each vortex from the center and the average radius of
the expanding condensate, by measuring them in the portrayed images
of the condensate after the time of flight.
During the expansion the
transverse lengths scale by the same factor \cite{Paris1,Paris2},
therefore the relation between the distance of a vortex from the
center and the average radius of the condensate ($r_{\perp 0}/R_\perp$)
will be the same as before the expansion.
From the transverse absorption images it can also be seen, that
vortex arrays have uniform spatial distribution and thus, vortices
are equidistant from the center as has been assumed in Figs.~2 and 3.
Then, with the values of $r_{\perp 0}/R_\perp$ extracted from \cite{Paris1}
by using Eq.~(\ref{Lz/Narray}) the following estimates for the
angular momentum per particle are obtained:
$<l_z>/\hbar= 1, 1.33, 1.36$ and $1.38$,
corresponding to the experimental configurations with $1,~2,~3$ and
$4$ vortices, respectively. These values of $<l_z>/\hbar$ are smaller than
the equilibrium values given in Fig.~3 which seems to indicate that the
relative radii of the vortex-arrays measured in \cite{Paris1} are larger than  
their calculated
equilibrium values in the rotating trap.
It would be nice to compare in detail the  values of $<l_z>/\hbar$ we
calculate from the images in \cite{Paris1} with those measured directly in
\cite{Paris2}. Unfortunately, however,
the experimental parameters  are slightly different in \cite{Paris1} and
\cite{Paris2} which makes a comparison difficult.
But it is worth stressing that the calculated result for the angular momentum 
 qualitatively agrees with the
experimental data in \cite{Paris2}:
after the first jump of $\hbar$ corresponding to the nucleation of
a singly quantized vortex centered in the trap,
the angular momentum increases
continuously when the number of vortices in
the array increases \cite{Butts99},
(which is equivalent to have increased the stirring frequency)
without presenting other jumps of order $\hbar$.

\section{Energy splitting}
The presence of a quantized vortex in a confined condensate breaks
time reversal symmetry: this produces a frequency shift of the modes
with azimuthal quantum number $\pm |m|$, which
in the absence of vortices are degenerated.
We want to calculate the frequency splitting of the low lying modes
in large systems due to an off-center vortex.

Theoretical calculations of the frequency shift produced by a
centered vortex have been performed within different approaches:
a sum rule approach \cite{sumrule},
a semiclassical approach based on a large $N$
expansion \cite{Sinha97}, a hydrodynamic approach \cite{Svidzinsky98},
and a full numerical solution of the linearized equations of motion
\cite{Dodd97}.

Current experiments in ENS \cite{Paris2}
have excited the two transverse
quadrupole modes $m=\pm 2$ of a quasi-cylindrically symmetric condensate.
Due to the presence of a vortex, there exists a lift of degeneracy
between the frequencies of these two quadrupole modes that causes a
precession of the system. Measuring the frequency of precession and from
the analytical expressions of the frequency shift \cite{sumrule} the
angular momentum of the system is inferred.

For simplicity, we will assume a large condensate trapped in a
cylindrically symmetric trap (\ref{Vtrap})
and we will study the energy splitting for the low-lying modes
due to an off-center vortex at a distance $r_0$ from the symmetry axis.
In the following we shall use the perturbative approach 
\cite{Svidzinsky98,Svidzinsky00}.
(The perturbative result for the frequency-splitting of the $m=\pm2$ modes
induced by
an off-centered vortex in a trap with different symmetry from the one  
considered here was reported in \cite{Svidzinsky01}).
Linearizing the Gross-Pitaevskii equation around the condensate 
using the decomposition (\ref{wf}) of $\Psi$ in condensate density
and phase, and assuming the Thomas-Fermi approximation
and the long-wavelength limit, it follows the coupled equations
\begin{eqnarray}
i\omega \,\delta\rho&&={\bbox \nabla}\!\cdot\!({\bf v}_0 \,\delta\rho)  
+\frac{\hbar}{M}\bbox{\nabla}
\cdot(\rho_0 {\bbox \nabla}\,\delta S)\label{a}\\
i\omega \,\delta S&&={\bf v}_0\cdot{\bbox \nabla}\,\delta S+
\frac{g}{\hbar}\,\delta\rho 
%i\omega \rho'&&=\bf{\nabla}\cdot (\bf{v}_0 \rho')  
%+\frac{\hbar}{M}\vec{\nabla}\cdot(\rho_0\bf{\nabla}\phi')\label{a}\\
%i\omega \phi'&&=\bf{v}_0\cdot\bf{\nabla}\phi'+\frac{g}{\hbar}\rho' \,.
\label{b}
\end{eqnarray}
where $\delta\rho$ and $\delta S$ are small deviations 
from the equilibrium values of the density and phase, respectively.
Solving (\ref{b}) for $\delta S$ to first
order in ${\bf v}_0$ and inserting in (\ref{a}) to eliminate $\delta S$,
yields the perturbed wave-equation
\begin{equation}
\left(\omega^2+\frac{g}{M}\bbox{\nabla}\!\cdot\rho_0
\bbox{\nabla}\right) \,\delta\rho =-i\,\omega\, {\bbox \nabla}
\!\cdot\!({\bf v}_0\,\delta\rho)+\frac{ig}{\omega M}{\bbox \nabla}\!
\cdot\! \Big( \rho_0{\bbox\nabla}({\bf v}_0
\!\cdot\!{\bbox\nabla}\delta\rho) \Big)\,.
%\left(\omega^2+\frac{g}{M}\bbox{\nabla}\!\cdot\!\rho_0({\bf r})
%\bbox{\nabla}\right) \,\delta\rho =-i\,\omega\, {\bbox \nabla}
%\!\cdot\!({\bf v}_0({\bf r})\,\delta\rho)+\frac{ig}{\omega M}{\bbox \nabla}\!
%\cdot\! \Big( \rho_0({\bf r}){\bbox\nabla}({\bf v}_0({\bf r})
%\!\cdot\!{\bbox\nabla}\delta\rho) \Big)\,.
\label{c}
\end{equation}
In first-order perturbation-theory the shift of the eigenvalue $\omega^2$ is  
obtained from the expectation-value of the perturbation, taken with the  
unperturbed solution of (\ref{c}), $\omega_0^2$, with ${\bf v}_0=0$:
\begin{equation}
\omega^2-\omega_0^2=-i\,\omega_0<{\bf v}_0\cdot {\bbox \nabla}>
+\frac{ig}{\omega_0M}
<({\bbox \nabla}\!\cdot\!\rho_0{\bbox \nabla})
({\bf v}_0\!\cdot\!{\bbox \nabla})>
\label{d}
\end{equation}
The expectation-value can be evaluated using the unperturbed Thomas-Fermi  
density (\ref{TFdensity})
%$\rho_0(\bf{r})\sim (1-r_\perp^2/R_\perp^2)$, 
if care is taken to interpret  
the radial part of the space-integral as a principal-value integral at  
$r_\perp=r_0$, because the actual condensate density at 
the position of the vortex
strictly vanishes.

Let us evaluate then the expectation-value for a single 
off-centered vortex (\ref{vimage}) in the particular case of surface
modes, for which \cite{Stringari96}
\begin{equation}
\delta \rho=\sqrt{\frac{m+1}{\pi}}
\frac{r_\perp^m }{R^{m+1}_\perp}\,e^{im\varphi}\,,\qquad
\omega_0=\sqrt{m}\,\omega_\perp
\end{equation}
where we take $m$ first as positive. We obtain after some calculation (and  
using to considerable advantage
the technique referred to in the preceding footnote) the following result
for the frequency shift, due to the off-centered vortex, exhibited by
the surface mode with positive $m$
\begin{equation}
\omega-\omega_0=\frac{(m+1)}{MR_\perp^2}\,
\left(1-\frac{r_0^{2m-2}}{R_\perp^{2m-2}}\right) \,,
\label{e}
\end{equation}
It turns out during the calculation that the mirror-vortex makes
actually no
contribution to the frequency-shift.
Because the velocity-field enters linearly, the frequency shift induced by  
several vortices is additive. Replacing $m$ by $-m$ is equivalent to inverting
${\bf v}_0$, i.e. $\omega-\omega_0$ simply changes sign.
The frequency splitting between the two frequencies (with $m$ and $-m$) 
is then
\begin{equation}
\omega_+-\omega_-=\frac{2 m}{MR_\perp^2}
\left(1-\frac{r_0^{2m-2}}{R_\perp^{2m-2}}\right).
\label{split}
\end{equation}
It can be seen from (\ref{e}) that there is no frequency shift for 
the dipole mode $m=1$,
at least not to first order in ${\bf v}_0$, i.e., the Kohn theorem, which  
predicts $\omega(m=1)=\omega_\perp$, is duely respected.

The splitting of the quadrupole excitations
($m=2$) has been measured in  
the ENS experiment \cite{Paris2}. For this case it follows from (\ref{split})
\begin{equation}
\omega_+-\omega_-=\frac{6}{MR_\perp^2}
\left[1-\left(\frac{r_0}{R_\perp}\right)^2\right] 
\label{f}
\end{equation}
From Eq.~(\ref{f}) one can see that there is a term in the
frequency shift that depends on the distance of the off-center
vortex from the center. It means that the precession of the
eigenaxis of the quadrupole mode,
$\dot{\theta}=(\omega_+-\omega_-)/2|m|$
that is measured in ENS experiment \cite{Paris2}, will be
slightly different for an off-center than for a centered vortex.
And thus, from this difference the distance of the off-centered vortex
can be inferred.

It is worth noting that Eq.~(\ref{f}) in the particular case
of a centered vortex ($r_0/R_\perp=0$) does not give the same dependence
on the trap and condensate parameters as the result obtained in
\cite{sumrule,Svidzinsky00} because the symmetry of the trap
considered in both calculations is different.

The preceding calculation can be extended to axial vortex-modes
\cite{lev}
travelling along the axis of the trap with wave-number $k$.
For small $k$ the unperturbed modes without vortex, to leading
order in $(k R_\perp)^2$, take the form
\begin{equation}
\delta\rho(r_\perp,\varphi,z)=\sqrt{\frac{m+1}{\pi}}\left[1+\frac{k^2}
{4(m+2)}\left(r_\perp^2-\frac{m+1}{m+2}\,R_\perp^2\right)\right]
\frac{r_\perp^m}{R^{m+1}_\perp}
\,e^{im\varphi}\,e^{\pm ikz}
\label{g}
\end{equation}
with frequencies
\begin{equation}
\omega_0=\sqrt{m}\,\omega_\perp\left(1+\frac{k^2R_\perp^2}{4m(m+2)}\right)\,.
\label{gg}
\end{equation}
Using (\ref{g}) and (\ref{gg})
in Eq.~(\ref{d}), we obtain after the evaluation of the expectation  
values to
leading order in
$(kR_\perp)^2$
\begin{eqnarray}
\omega_+-\omega_- = (\omega_+-\omega_-)_0
+ &&\frac{\hbar k^2}{2Mm(m+2)^2}\Big[m^2+m+2+4m\left(\frac{r_0}
{R_\perp}\right)^{2m+2}\nonumber\\
&&-2(m+1)^2(m+2)\left(\frac{r_0}{R_\perp}\right)^{2m}
+(m+1)(2m^2+3m+2)\left(\frac{r_0}{R_\perp}\right)^{2m-2}\Big]
\label{h}
\end{eqnarray}
where $(\omega_+-\omega_-)_0$ is the result following from (\ref{d})
to which (\ref{h}) reduces for $k=0$.
For the special case of a centered vortex $r_0=0$ and
taking $m=1$
the result (\ref{h}) reduces to a result derived previously
in \cite{Svidzinsky98}.

Let us now compare these results with those following from an application of  
the sum-rule approach, developed for this problem in \cite{sumrule}. As is
explained there, the sum-rule approach is based on the assumption that the  
expectation-values of the commutators of certain adequately chosen excitation  
operators, taken in the ground-state containing the vortex, will be exhausted
by the two modes $\omega_+$, $\omega_-$ whose splitting is to be calculated.
As has been shown in \cite{sumrule}, this approach works
very well (i.e., its basic assumption is satisfied to the required accuracy)
for the axially symmetric case of a centered vortex, where in the Thomas-Fermi
and long-wavelength limit the result can be checked by perturbation-theory
\cite{Svidzinsky00}. However, we shall now see that for the case of  
off-centered vortices the sum-rule approach fails, because the
result of first-order perturbation-theory is not reproduced.

First, we briefly recall the sum rule approach.
Let $F_\pm= \sum_{j=1}^N f_\pm({\bf r}_j)$ be the
mutually adjoint operators, carrying
opposite angular momentum, which excite the collective modes $\pm |m|$,
respectively.
% and satisfy $F^{\dagger}_{+}=F_{-}$.
Let $m_p^{\pm}$ be the $p$-energy weighted moments of the dynamic
structure factor associated to the excitation operators $F_{\pm}$,
which can be written as expectation values of commutators between
the many-body Hamiltonian of the system ($H$) and $F_{\pm}$:
\begin{equation}
m_1^{+}=<\!0 |\, [F_{-},[H,F_{+}]]\, | 0\!> =
         \frac{N \hbar^2}{M} <\!0|\,|{\bf \nabla} f_{+}|^2 |0\! >
\label{m1+}
\end{equation}
\begin{equation}
m_2^{-}=<\!0 |\,[[F_{-},H],[H,F_{+}]]\,| 0\!> =
         N <\!0|\,[j_{-},j_{+}]\,|0\!> \,,
\label{m2-}
\end{equation}
where $|0\!>$ may be a vortex-free or a vortex state, and
$j_{\pm}=(\hbar/m) {\bbox\nabla}f_{\pm} \cdot {\bf p}$.
Assuming that the moments are exhausted by the modes $\pm |m|$,
the shift of the collective frequencies in the TF limit
can be calculated as
\begin{equation}
\hbar (\omega_{+} - \omega_{-}) = m_2^{-}/m_1^{+} \,.
\label{shift}
\end{equation}

Applying this prescription let us start with the transverse quadrupole modes
$m=\pm 2$. The excitation
operators carrying angular momentum $m=\pm2$, are
\begin{equation}
f_{\pm}= (x \pm i y)^2 \,.
\label{fm2}
\end{equation}
Evaluating the moments Eqs.~(\ref{m1+}) and (\ref{m2-}), and
from (\ref{shift}) the following result for the frequency shift
for the $m=\pm2$ modes is obtained \cite{sumrule}:
\begin{equation}
\omega_{+}-\omega_{-}= \frac{2}{M} \frac{< l_z >}{<r_\perp^2>}\,.
\label{shiftm2}
\end{equation}
Neglecting the microscopic details of the vortex core structure
in the density profile,
the square radii of the condensate can be evaluated using the TF
approximation \cite{Trento}.
This prescription can be applied to a centered- or
non-centered vortex. What changes in each of these configurations
 is
the averaged value of the axial component of the angular momentum
$<l_z>$.
Using the results obtained in the previous Section,
 Eq.~(\ref{Lz/N})
the sum-rule result for the frequency-splitting due to an off-center vortex
at a distance $r_{0}$ from the symmetry axis, is
\begin{equation}
%\omega_{+}-\omega_{-}= \frac{3}{2} \frac{\omega_\perp \kappa}
%%(a N_\perp)^{-1/2}\left(1-\left(\frac{r_0}{R_\perp}\right)^2\right)^2
%{\sqrt{a N_\perp}}\left[1-\left(\frac{r_0}{R_\perp}\right)^2\right]^2\,,
\omega_{+}-\omega_{-}= \frac{6}{M R_\perp^2}
\left[1-\left(\frac{r_0}{R_\perp}\right)^2\right]^2\,,
\label{shift2}
\end{equation}
where we recall that the square radius in the transverse direction
of a cylindrical condensate in the TF limit is:
\begin{equation}
%<r_\perp^2>=\frac{R_\perp^2}{3}=\frac{4}{3}\,\sqrt{a N_\perp}
%   \, a_\perp^2
<r_\perp^2>=\frac{R_\perp^2}{3}=\frac{4}{3}\,(a N_\perp)^{1/2}
       a_\perp^2 \,.
\label{r2}
\end{equation}
The splitting (\ref{shift2}) differs from the perturbative result 
(\ref{f}) by the factor
$[1-(r_0/R_\perp)^2]$ and therefore systematically
underestimates the frequency-shift, except for a centered vortex
$r_0=0$.

Let us try the sum-rule approach also on the calculation
of the splitting of axial helical vortex modes
\cite{lev} with low multi-polarity and small wave-number $k$.
The excitation operators of the axial helical vortex modes with
wave number $k$ and angular distortion $m=\pm1$ and $m=\pm2$
are
\begin{equation}
f_{\pm}= (x \pm i y)\, e^{\pm i k z} \,,
\label{fm1helical}
\end{equation}
%for $m=\pm1$, and for $m=\pm2$
\begin{equation}
f_{\pm}= (x \pm i y)^2 \, e^{\pm i k z} \,,
\label{fm2helical}
\end{equation}
respectively.
The calculation of the corresponding moments leads to the following
frequency shifts. For the $m=\pm1$ helical vortex mode:
\begin{equation}
\omega_{+}-\omega_{-}= \frac{k^2}{M}< l_z>
             \label{shiftm1helical}
\end{equation}
and for
 $m=\pm2$
\begin{equation}
\omega_{+}-\omega_{-} =
%         \frac{2 < l_z>}{M <r_\perp^2>}
%        \left[1+\frac{k^2}{4}<r_\perp^2>
%         \! \left(1-\frac{<r_\perp^4>}
%        {2<r_\perp^2>^2}\right)\right] \nonumber \\
   \frac{2 < l_z>}{M <r_\perp^2>}
     \left[1+\frac{k^2}{16}<r_\perp^2>\right]\,,
\label{shiftm2helical}
\end{equation}
where we have used
$<r_\perp^4>=3/2\,<r_\perp^2>^2$.
Both expressions agree with the perturbative result for centered vortices
$r_0=0$, but fail again to reproduce the correct first-order results
for the off-center vortices.

\section{Summary}
We have calculated the angular momentum of a large cylindrical
condensate in the presence of an off-axis vortex line and also
vortex arrays, within the Thomas Fermi approximation. In addition,
the interaction energy of vortex lines in the spatially inhomogeneous
condensate has been derived for the same case, including the effects
of the image vortices created by the boundary condition of vanishing normal
velocity. It is interesting that the image vortices do not contribute to the  
angular momentum even though they strongly modify the velocity-field around
any off-center vortex-line. The kinetic energy, on the other hand, is
influenced by the presence of the image-vortices.

We have seen that the contribution of a
given vortex to the angular momentum per particle decreases
as it moves away from the center of the system,
which leads to a smooth increase of the angular momentum as a new
vortex appears at a large distance from the axis of rotation.
We have obtained an estimate of the angular momentum corresponding
to different experimentally observed vortex configurations
\cite{Paris1}, that qualitatively agree with the continuous
behavior measured from experimental data \cite{Paris2}.
Finally, we have studied the frequency splitting of low lying
collective modes due to the presence of an off-centered vortex,
including axial vortex-waves,
using the perturbative approach \cite{Svidzinsky00}
and also the sum rule approach \cite{sumrule}. A comparison of the results
shows, that the simple extension of the sum-rule approach to off-axis
vortices does actually not work, i.e., the sum-rules are in this case
not exhausted by the two nearly degenerate modes of interest.

\section{Acknowledgments}
We thank A. Csord\'as for useful discussions.
M.G. thanks the ESF for financial support
and the Theoretische Physik group
%Fachbereich Physik of Universit\"at
at the Universit\"at GH Essen for the warm hospitality.
We would like to thank a referee for alerting us to the
doubtfulness of the sum-rule approach
in the case of off-axis vortices.

{\it Note added:}
After completion of this work, we have received a preprint
\cite{Garcia} reporting numerical calculations of critical
frequencies and angular momentum of vortex configurations with different
vorticity in an elongated trap. The vortices numerically obtained in
\cite{Garcia} are not rectilinear but with deformed shape and the
longitudinal deformation of these vortices is responsible for the
almost continuous increase of the angular momentum with respect to the
angular frequency.

\begin{figure}
%\centerline{\epsfig{figure=fig1.ps,width=7.5cm,angle=-90}}
\caption{
Dimensionless energy of symmetrical
vortex-arrays of 1,~2,~3,~4 singly quantized vortices
(from bottom curve to top curve,
respectively) as a function of the rescaled distance of the vortex lines
from the $z$-axis. The condensate has
$N=2.5 \times 10^5$ atoms of $~^{87}$Rb with
$R_\perp = 3.8 \mu m$, which correspond to the experimental
parameters of Ref.~\protect\cite{Paris2} }
\label{fig1}
\end{figure}
\begin{figure}
%\centerline{\epsfig{figure=fig2.ps,width=7.5cm,angle=-90}}
\caption{
Dimensionless energy of symmetrical
vortex-arrays of 1,~2,~3,~4 singly quantized vortices
(from bottom curve to top curve,
respectively) in a trap rotating with frequency $\Omega/2\pi=40$ Hz
as a function of the rescaled distance of the vortex lines
from the $z$-axis. The condensate has
$N=2.5 \times 10^5$ atoms of $~^{87}$Rb with
$R_\perp = 3.8 \mu m$, which correspond to the experimental
parameters of Ref.~\protect\cite{Paris2} }
\label{fig2}
\end{figure}
\begin{figure}
%\centerline{\epsfig{figure=fig3.ps,width=7.5cm,angle=-90}}
\caption{
Average angular momentum per particle of symmetrical arrays of 1,~2,~3,~4
vortices (from bottom curve to top curve, respectively) in relative equilibrium
in a trap rotating with frequency $\Omega/2\pi$ as a function of $\Omega$.
The values of $M$ and $R_\perp$ are the same as in Figs.~1,2}
\label{fig3}
\end{figure}

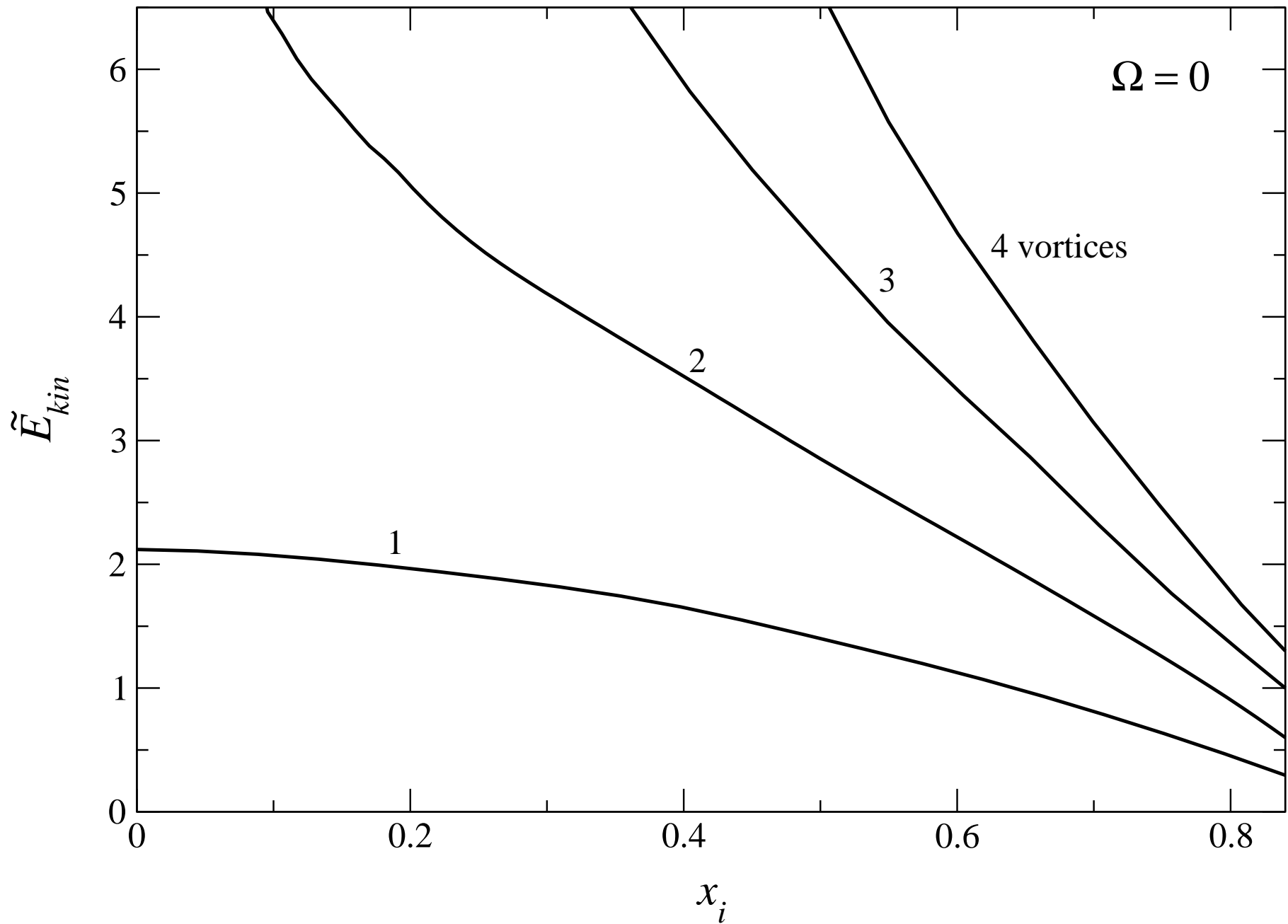
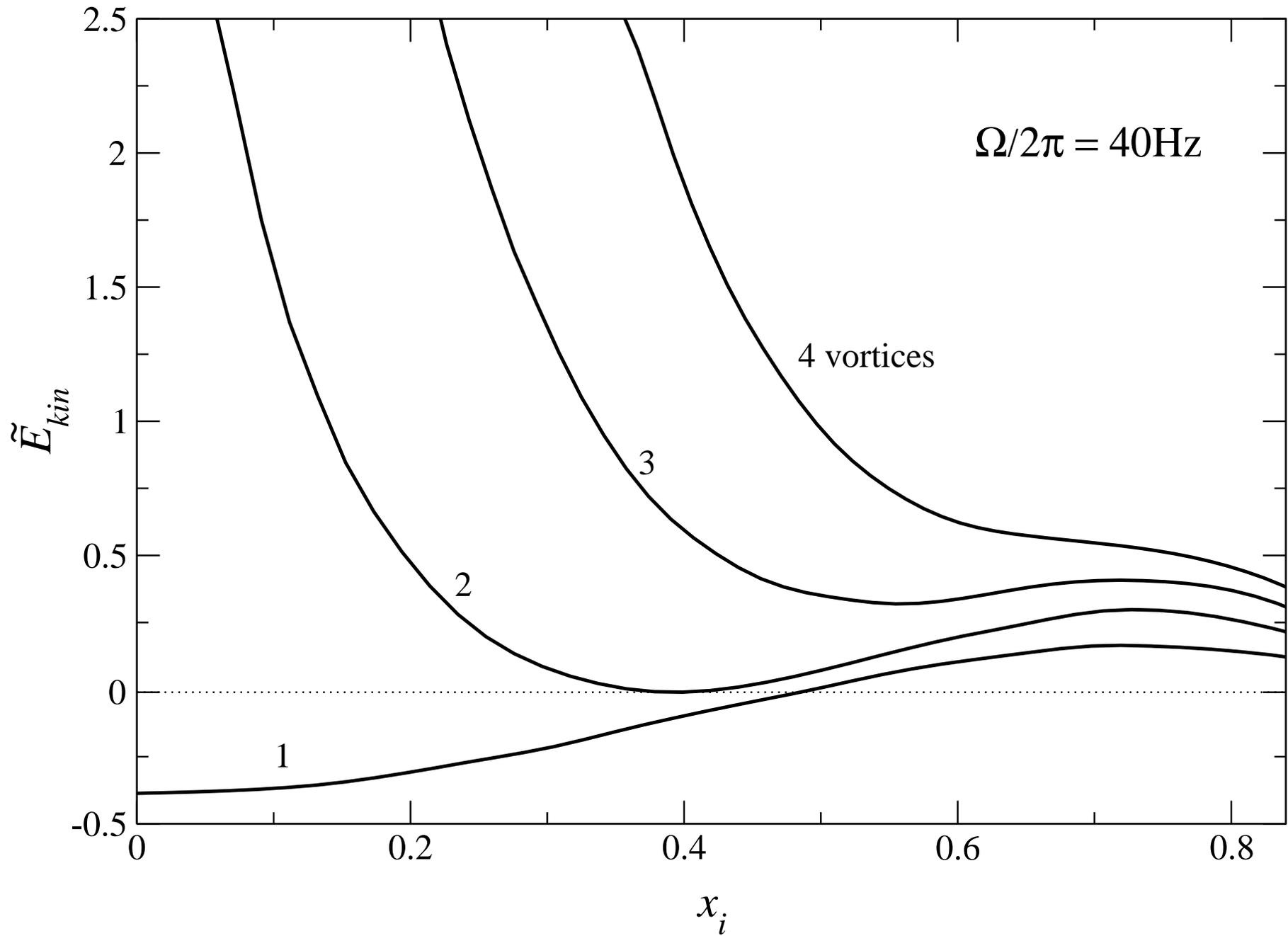
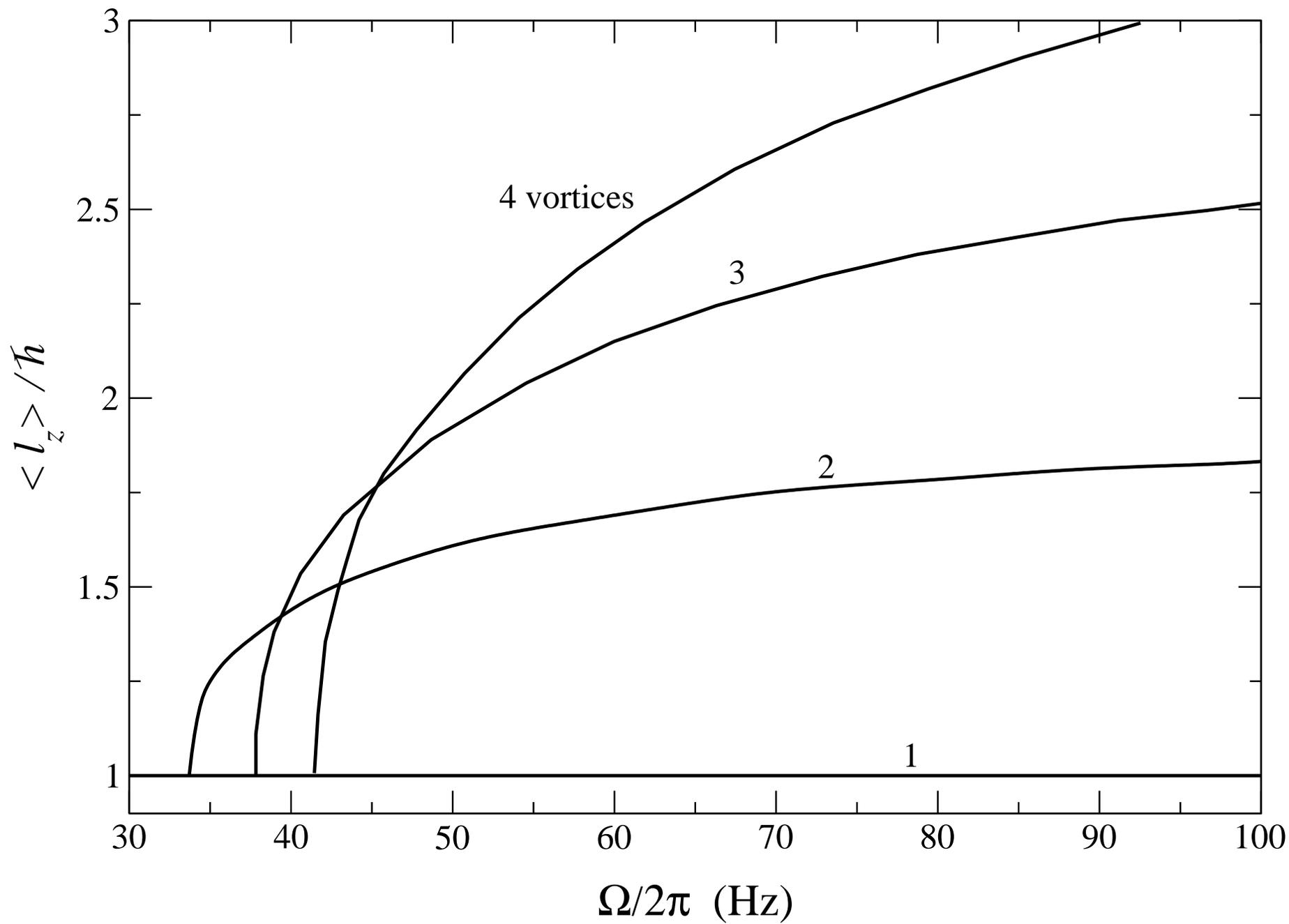
\end{document}